\documentclass[conference]{IEEEtran}

\usepackage{graphicx}
\usepackage{color}
\usepackage[caption=false,font=footnotesize]{subfig}
\usepackage{cite}

\usepackage{amsmath}
\usepackage{ascmac}
\usepackage{amssymb}
\usepackage{amsfonts}
\usepackage{mathtools}
\usepackage{bm,bbm}
\usepackage{booktabs}

\usepackage{algorithm,algpseudocode}

\graphicspath{{./_figs/}}

\renewcommand{\figurename}{Fig.}
\renewcommand{\tablename}{TABLE}
\newcommand{\sectionname}{Section}

\newcommand{\noindentbf}[1]{\noindent\textbf{#1}\quad}

\newcommand{\vspacelb}{\vspace{+0.3em}\\}

\begin{document}

\title{
    Respiratory Rate Estimation \\ Based on WiFi Frame Capture
}

\author{
    \IEEEauthorblockN{
        \normalsize Takamochi Kanda\IEEEauthorrefmark{1},
        \normalsize Takashi Sato\IEEEauthorrefmark{1}\IEEEauthorrefmark{2},
        \normalsize Hiromitsu Awano\IEEEauthorrefmark{1}\IEEEauthorrefmark{3},
        \normalsize Sota Kondo\IEEEauthorrefmark{1}, and
        \normalsize Koji Yamamoto\IEEEauthorrefmark{1}\IEEEauthorrefmark{4}
    }
    \vspace{+0.5pt}
    \IEEEauthorblockA{
        \IEEEauthorrefmark{1}\small Graduate School of Informatics, Kyoto University,
        Yoshida-honmachi, Sakyo-ku, Kyoto, 606-8501, Japan \\
        \IEEEauthorrefmark{2}\small takashi@i.kyoto-u.ac.jp,
        \IEEEauthorrefmark{3}\small awano@i.kyoto-u.ac.jp,
        \IEEEauthorrefmark{4}\small kyamamot@i.kyoto-u.ac.jp
    }
}

\maketitle
\begin{abstract}

    This paper presents a method
    that estimates the respiratory rate based on the
    frame capturing of wireless local area networks.
    The method uses beamforming feedback matrices (BFMs)
    contained in the captured frames,
    which is a rotation matrix of channel state information (CSI).
    BFMs are transmitted unencrypted and easily obtained using frame capturing,
    requiring no specific firmware or WiFi chipsets,
    unlike the methods that use CSI.
    Such properties of BFMs allow us to apply frame capturing
    to various sensing tasks, e.g., vital sensing.
    In the proposed method,
    principal component analysis is applied to BFMs
    to isolate the effect of the chest movement of the subject,
    and then, discrete Fourier transform is performed
    to extract respiratory rates in a frequency domain.
    Experimental evaluation results confirm that
    the frame-capture-based respiratory rate estimation can achieve
    estimation error lower than 3.2 breaths/minute.

\end{abstract}
\IEEEpeerreviewmaketitle

\section{Introduction}
\label{sec:introduction}

Monitoring vital signs, e.g., heart rate or
respiratory rate, plays an important role in medical
care, particularly after the global outbreak of COVID-19.
Patients with serious symptoms need to constantly monitor their vital signs because their
symptoms may suddenly worsen as observed in COVID-19 cases~\cite{wan2020clinical} or
acute respiratory distress syndrome (ARDS)~\cite{force2012acute}.
In recent times, the continuous sensing of biological signals
relies on wired devices,
which have been a burden for patients wearing them.
One solution to realize contact-less vital sensing uses millimeter-wave radar~\cite{antide2020comparative}.
However, it requires radar transmitters and receivers, which significantly increases the initial cost.

To alleviate this problem, vital sensing using WiFi devices
has been installed in several places and is attracting a lot of attention.
Conventionally, WiFi-based vital sensing
relies on the received signal strength indicator (RSSI),
which represents the received signal strength measured for each frame.
However, owing to the low frequency and time resolution of RSSI, in recent years,
the methods that use channel state information (CSI) has been proposed~\cite{he2020wifi,ma2019wifi}.
The CSI carries channel information for each subcarrier as a multidimensional matrix.
Owing to fine-grained channel information provided by CSI,
respiratory rate~\cite{liu2014wi,liu2015contactless}
and heart rate~\cite{liu2015tracking} can be accurately measured.
Although the CSI-based approaches require no additional devices,
retrieving CSI information requires firmware modifications or dedicated WiFi chipsets,
which limit the applicability of CSI-based vital sensing.

Another promising
technique is a frame-capture-based approach that uses
beamforming feedback matrices (BFMs)
defined in IEEE 802.11ac/ax standard~\cite{6687187,9442429}.
For example, frame-capture-based indoor localization and outdoor human detection
have been implemented using machine learning techniques~\cite{fukushima2019evaluating, miyazaki2019initial, takahashi2019dnn}.
Beamforming feedback frames that deliver BFMs can be
easily captured by third parties because they are unencrypted.
Moreover, capturing beamforming feedback frames does not require any special firmware or WiFi chipsets, which
would have been necessary to obtain CSI.
These facts can extend the applicability of frame capturing to various sensing tasks.

In this paper, we propose a respiratory rate estimation method and
demonstrate the feasibility of vital sign monitoring based on frame capturing.
Assuming that the chest movement owing to respiration is the dominant cause of
signal fluctuation, first,
principal component analysis (PCA) is applied to BFMs.
Then, a discrete Fourier transform (DFT) is performed
to extract the respiratory rate in the frequency domain.

The main contributions of this paper are summarized as follows:
\begin{itemize}
    \item We propose a frame-capture-based respiratory rate estimation,
    which extends the applicability of WiFi-based vital sensing.
    The proposed method does not rely on any learning-based algorithms, and hence, it
    can be applied to various environments without incurring any cost related to learning.
    More specifically, we isolate the effect of
    chest fluctuation of the subject from other channel information by applying PCA to the captured BFMs
    and perform DFT on the first principal component to extract the periodic change
    in the signal caused by human breathing.
    \item We show the feasibility of accurate respiratory rate estimation based on
    frame capturing without specific firmware modifications or WiFi chipsets
    through experiments.
    The experimental results confirm that
    the proposed method can achieve
    estimation error lower than $3.2\,\mathrm{breaths/minute}$.
\end{itemize}

The rest of this paper is organized as follows:
\sectionname~\ref{sec:system_model} describes a system model and
BFMs in the captured frames.
\sectionname~\ref{sec:proposed_method} explains the proposed frame-capture-based respiratory rate estimation method.
\sectionname~\ref{sec:evaluation} describes experimental evaluation results and discusses the results.
\sectionname~\ref{sec:conclusion} concludes the paper.

\section{System Model}
\label{sec:system_model}

\begin{figure}[t]
    \centering
    \includegraphics[width=\columnwidth]{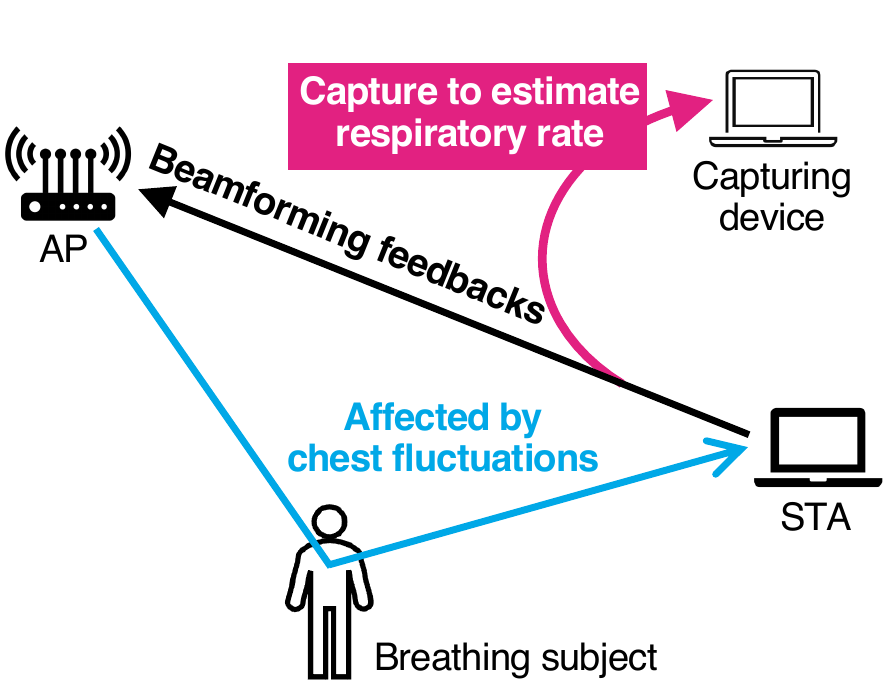}
    \caption{
        Frame-capture-based respiratory estimation system.
        The STA transmits beamforming feedback frames to the AP,
        and the capturing device captures these frames to
        estimate the subject's respiratory rate.
    }
    \label{fig:system_model}
\end{figure}

\figurename~\ref{fig:system_model} shows the proposed
respiratory rate estimation system based on frame capturing.
The system comprises an access point (AP) with $N_{\mathrm{AP}}$ antennas,
a station (STA) with $N_{\mathrm{STA}}$ antennas, and a capturing device,
where we assume $N_{\mathrm{AP}} \geq N_{\mathrm{STA}}$.
One subject breathes at rest in the vicinity of the AP and STA.
The STA is associated with the AP, and the AP periodically transmits data frames to the STA.
The capturing device captures the transmitted frames by the STA.

We exploit the beamforming feedback frames transmitted periodically
from the STA to the AP according to the beamforming feedback procedure
defined in IEEE 802.11ac/ax for the respiratory rate estimation.
The beamforming feedback frames contain the elements of the BFM,
$\bm{V}\in\mathbb{C}^{N_{\mathrm{AP}}\times N_{\mathrm{STA}}}$,
which is defined as the right singular matrix of the CSI matrix
$\bm{H}\in\mathbb{C}^{N_{\mathrm{STA}}\times N_{\mathrm{AP}}}$:
\begin{align}
    \bm{H} &= \bm{U}\bm{\varSigma}\bm{V}^{\mathrm{H}}, \\
    \bm{y} &= \bm{H}\bm{x},
\end{align}
where $\bm{y}\in\mathbb{C}^{N_{\mathrm{STA}}}$ and $\bm{x}\in\mathbb{C}^{N_{\mathrm{AP}}}$
are the receive and transmit signal vectors, respectively,
$\bm{U}\in\mathbb{C}^{N_{\mathrm{STA}}\times N_{\mathrm{STA}}}$
is the left singular matrix,
$\bm{\varSigma}\in\mathbb{C}^{N_{\mathrm{STA}}\times N_{\mathrm{STA}}}$ is the diagonal matrix
whose entries are singular values,
and the superscript $\mathrm{H}$ denotes the Hermitian transpose.

As the elements of BFMs are transmitted in a compressed
form,
decompression is firstly performed
before applying our respiratory rate estimation procedure.
More concretely,
BFM $\bm{V}$ is expressed by
$\phi_{i, j}\in[0, 2\pi],\ \psi_{i, j}\in[0, \pi/2]$
based on Givens rotation
as referred to in IEEE 802.11ac standard \cite{6687187}:
\begin{align}
    \bm{V} = \left[
    \prod_{i=1}^{\min\{N_{\mathrm{AP}}, N_{\mathrm{STA}} - 1\}} \!\!\!\!\! \bm{D}_{i}\left(
    \prod_{l=i+1}^{{N_{\mathrm{STA}}}}\bm{G}_{l, i}(\psi_{l, i})^{\mathrm{T}}
    \right)
    \right]\tilde{\bm{I}}_{N_{\mathrm{AP}}\times N_{\mathrm{STA}}},
\end{align}
where
\begin{align}
    \bm{D}_{i} \coloneqq
    \begin{bmatrix}
        \bm{I}_{i-1} & 0                         & \cdots & \cdots                                   & 0 \\
        0            & e^{\mathrm{j}\phi_{i, i}} & 0      & \cdots                                   & 0 \\
        \vdots       & 0                         & \ddots & 0                                        & 0 \\
        \vdots       & \vdots                    & 0      & e^{\mathrm{j}\phi_{N_{\mathrm{AP}}-1, i}} & 0 \\
        0            & 0                         & 0      & 0                                        & 1
    \end{bmatrix}
\end{align}
is the diagonal matrix and
\begin{align}
    \bm{G}_{l, i}(\psi)
    \coloneqq
    \begin{bmatrix}
        \bm{I}_{i-1} & 0           & 0              & 0          & 0                         \\
        0            & \cos\psi  & 0              & \sin\psi & 0                         \\
        0            & 0           & \bm{I}_{l-i-1} & 0          & 0                         \\
        0            & -\sin\psi & 0              & \cos\psi & 0                         \\
        0            & 0           & 0              & 0          & \bm{I}_{N_{\mathrm{AP}}-l}
    \end{bmatrix}
\end{align}
is the Givens rotation matrix.
In these equations, $\tilde{\bm{I}}$ denotes the rectangular
version of the identity matrix $\bm{I}$,
i.e., main diagonal is $1$ and other elements are represented by $0$.
In addition, ${\phi_{i, j}}$ and ${\psi_{i, j}}$ are respectively quantized to
$b_{\phi}$ and $b_{\psi}$ bits as
\begin{align}
    \phi_{i, j} & = \frac{k_{\phi_{i, j}}\pi}{2^{b_{\phi}-1}} + \frac{\pi}{2^{b_{\phi}}}, \ \
    \psi_{i, j} = \frac{k_{\psi_{i, j}}\pi}{2^{b_{\psi}+1}} + \frac{\pi}{2^{b_{\psi}+2}},
\end{align}
which are termed compressed BFMs.

\section{Frame-Capture-Based Respiratory Rate Estimation}
\label{sec:proposed_method}

\begin{figure}[t]
    \centering
    \includegraphics[width=\columnwidth]{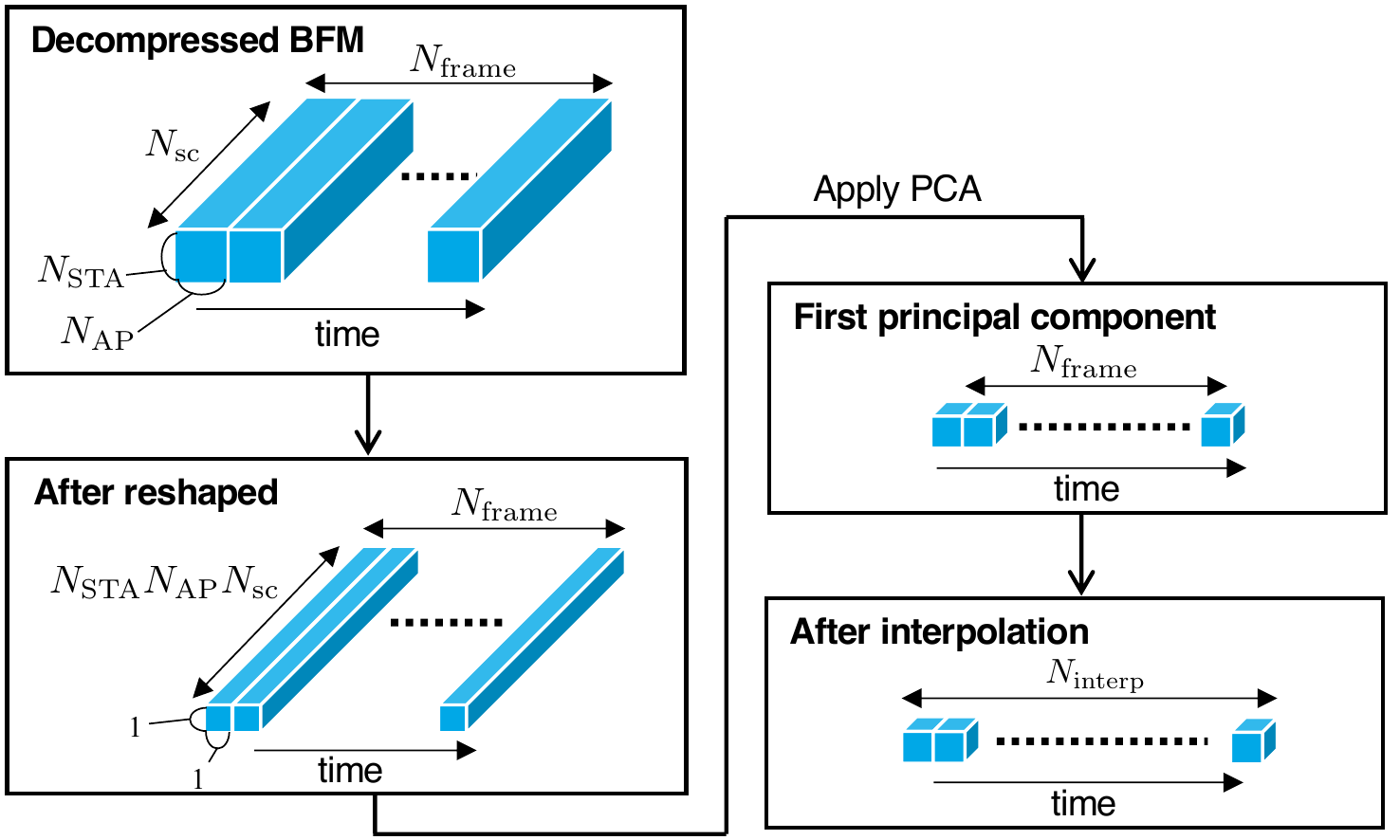}
    \caption{
        Data arrangements during the preprocessing.
        The decompressed BFM data are preprocessed into a one-dimensional
        vector along time.
    }
    \label{fig:data_shape}
\end{figure}

We describe the overview of
our frame-capture-based respiratory rate
estimation method below.
First, we isolate the effect of the chest fluctuation of the subject
from other channel information by applying PCA to the decompressed data.
Second, DFT is performed on the first principal component to extract
the periodic change in the signal caused by human breathing, and finally,
the respiratory rate is estimated by detecting the peak in the DFT signal.
The details of each step are as follows:
\vspacelb
\noindentbf{Preprocessing.}
The decompressed BFM data is processed by a time window.
Respiratory rate is estimated in real time by moving the time window
in a constant interval so that it overlaps the adjacent time window.
To apply PCA, the captured data is arranged two-dimensionally as a matrix of
size $N_{\mathrm{frame}} \times N_{\mathrm{STA}}N_{\mathrm{AP}}N_{\mathrm{sc}}$ as shown in Fig.~\ref{fig:data_shape},
where $N_{\mathrm{frame}}$ is the number of frames in the time window and $N_{\mathrm{sc}}$
is the number of subcarriers.
We use the amplitude of decompressed BFMs rather than phase because
the preliminary experiments confirmed that
the amplitude of BFM synchronizes better with the subject's chest movement than the phase.

Second, we apply PCA to the data to separate the effect of
chest movement from other channel information.
We use the first principal component that most reflects the chest movement.
Finally, we linearly interpolate the data
to equalize the time intervals before performing DFT
in the subsequent steps.
\figurename~\ref{fig:data_shape} shows changes in data shape based on the preprocessing procedure,
wherein $N_{\mathrm{interp}}$ denotes the number of interpolated points.
\vspacelb
\noindentbf{Respiratory Rate Estimation.}
We then extract the respiratory rate as the peak frequency
of the DFT output.
In order to avoid false peak detection,
first, we use a band-pass filter to limit the output of the DFT to the frequency range of the human respiration rate.
We then calculate the ratio of the peak value to the average value of the DFT output.
If the ratio is less than or equal to threshold $\theta$, we estimate it as 0,
meaning that there is no respiration or the subject is not breathing.
On the other hand, the ratio higher than $\theta$ indicates that respiration is detected.
During the period the respiration is detected,
the respiratory frequency is estimated as the frequency that gives the peak in the DFT signal.

\begin{figure}[t]
    \vspace{-3mm}
    \makebox[\columnwidth]{
        \begin{minipage}{\columnwidth}
            \begin{algorithm}[H]
                \caption{Respiration detection and estimation}
                \label{alg:proposed_method}
                \begin{algorithmic}[1]
                    \Require{Preprocessed data $\bm{x}$}
                    \State{Perform DFT on $\bm{x}$, and store the output in $\bm{y}$.}
                    \State{\Comment{Filter by a band-pass filter $L[\cdot]$ in TABLE~\ref{tbl:environment_settings}}}
                    \State{$\bm{y} \gets L[\bm{y}]$}
                    \If{$\mathrm{max}(\bm{y}) / \mathrm{mean}(\bm{y}) < \theta$}
                    \Comment{No respiration}
                    \State{$\mathrm{estimation}\gets 0.0$}
                    \Else{} \Comment{Respiration is detected}
                    \State{$\mathrm{estimation}\gets \mathrm{argmax}(\bm{y})$}
                    \EndIf{}
                    \Ensure{$\mathrm{estimation}$}
                \end{algorithmic}
            \end{algorithm}
        \end{minipage}
    }
\end{figure}

\section{Experimental Evaluation}
\label{sec:evaluation}

In this section, we evaluate the proposed respiratory rate estimation method.
First, experimental settings, including parameter and implementation settings, are provided.
Second, we show experimental results and some discussions about our preprocessing, DFT output,
and estimation.
To evaluate the performance of the
estimation, we validate the estimation error,
which is computed by root mean squared error (RMSE) between the estimation and ground-truth in the time domain.

\subsection{Experimental Setup}

We perform our experiments in a room of a concrete building using IEEE 802.11ax WLAN system.
The detailed parameter settings of the experiments are
listed in \tablename~\ref{tbl:environment_settings}.
The frequency band is $5.2\,\mathrm{GHz}$, and
the bandwidth is $80\,\mathrm{MHz}$, where the subcarrier interval for beamforming feedback in the frequency domain is
$312.5\,\mathrm{kHz}$, and $N_{\mathrm{sc}}=250$ BFMs are returned to the AP.
Throughout the experiments,
we use off-the-shelf WiFi access point WXR-5700AX7S (BUFFALO) as the AP and STA,
and Jetson Nano (NVIDIA) with Intel AX200 802.11ax WiFi card as the capturing device.
The AP transmits data frames to the STA using the \texttt{iperf3} command.
The beamforming feedback frames are captured using the capturing device with the \texttt{tshark} command,
which creates pcap network capture files.

\begin{figure}[t]
    \centering
    \subfloat[Photograph of an experiment.]{\includegraphics[width=\columnwidth]{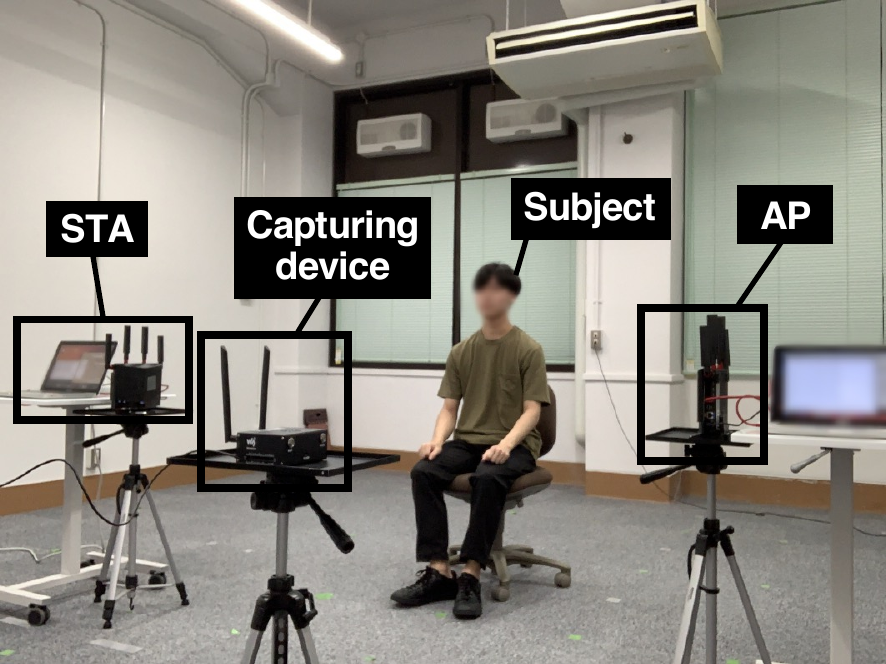}} \\
    \subfloat[Arrangement of devices.]{\includegraphics[width=\columnwidth]{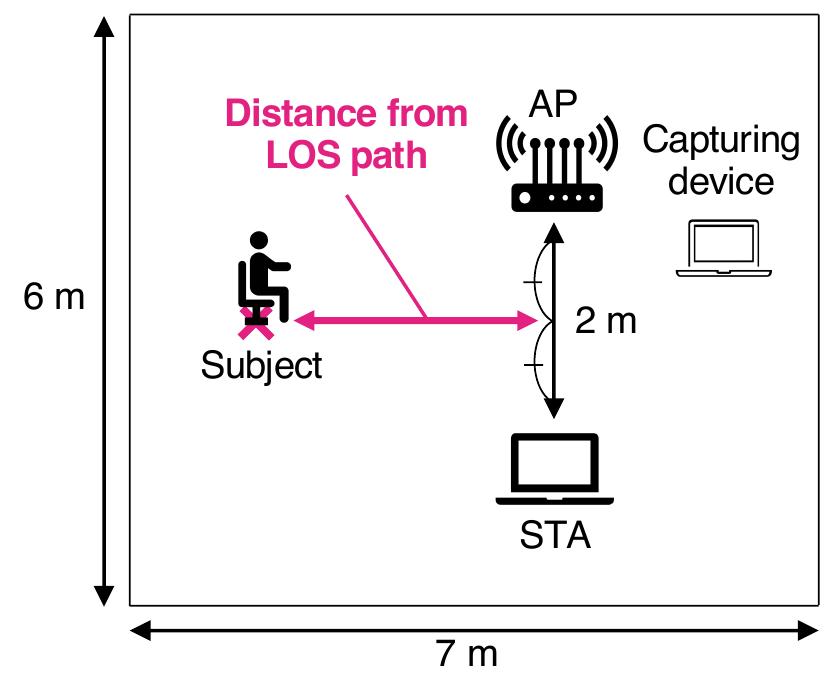}}
    \caption{
        Experimental setup.
        The distance between the AP and STA is $2\,\mathrm{m}$,
        and one subject is sitting on a chair.
    }
    \label{fig:experimental_environtment}
\end{figure}

\figurename~\ref{fig:experimental_environtment} depicts the experimental setting.
The distance between the AP and STA is $2\,\mathrm{m}$,
and the subject sits deep on a chair.
The room is furnished with desks, chairs, shelves,
air conditioners, and ventilation fans.
The subject is a 22-year-old-male; he is $1.68\,\mathrm{m}$ tall and weighs $56\,\mathrm{kg}$.
The height of the chest, the antennas of AP and STA are arranged at the same height.

We experimented with the subject at three positions,
where the distances from the middle of the line of sight (LOS) path
are $1$, $2$, and $3\,\mathrm{m}$.
At each positions, the subject breathes at
constant rate of $0$, $15$, and $20\,\mathrm{breaths/minute}$
in sync with a metronome throughout $300\,\mathrm{s}$.
The respiratory rate is estimated
based on the distance from each position and breathing rate, shifting the time window of $60\,\mathrm{s}$ overlapping $1\,\mathrm{s}$ with adjacent window.
The breathing rate $0\,\mathrm{breaths/minute}$ is the state in which the subject is holding breath.
As it is impossible for the subject to hold his breath throughout the experiment,
we give the subject time to breath in between,
which does not significantly affect the respiratory rate estimation because the estimation does not
depend on a slight change in the posture of the subject.

\begin{table}[t]
    \centering
    \caption{Experimental Parameters}
    \begin{tabular}[t]{cc}
        \toprule
        Parameters                                        & Values                     \\
        \midrule
        Average beamforming feedback interval             & $0.20\,\mathrm{s}$               \\
        Carrier frequency                                         & $5.2\,\mathrm{GHz}$              \\
        Bandwidth                                         & $80\,\mathrm{MHz}$               \\
        Number of AP's antennas $N_{\mathrm{AP}}$          & $4$                              \\
        Number of STA's antennas $N_{\mathrm{STA}}$         & $4$                              \\
        Number of subcarriers with BFMs $N_{\mathrm{sc}}$ & $250$                            \\
        \begin{tabular}{c}
            Capturing time at each\vspace{-1.5pt}\\positions and breathing rate
        \end{tabular}
        & $300\,\mathrm{s}$ \\
        Time window size                                       & $60\,\mathrm{s}$                 \\
        Interval of linear interpolation           & $0.1\,\mathrm{s}$                \\
        Band-pass filter $L$ range                    & $[10, 50]\,\mathrm{breaths/minute}$ \\
        Detection threshold $\theta$                      & $5.0$                            \\
        \bottomrule
    \end{tabular}
    \label{tbl:environment_settings}
\end{table}

\subsection{Results}

\begin{figure}[t]
    \centering
    \subfloat[Principal components.]{\includegraphics[width=\columnwidth]{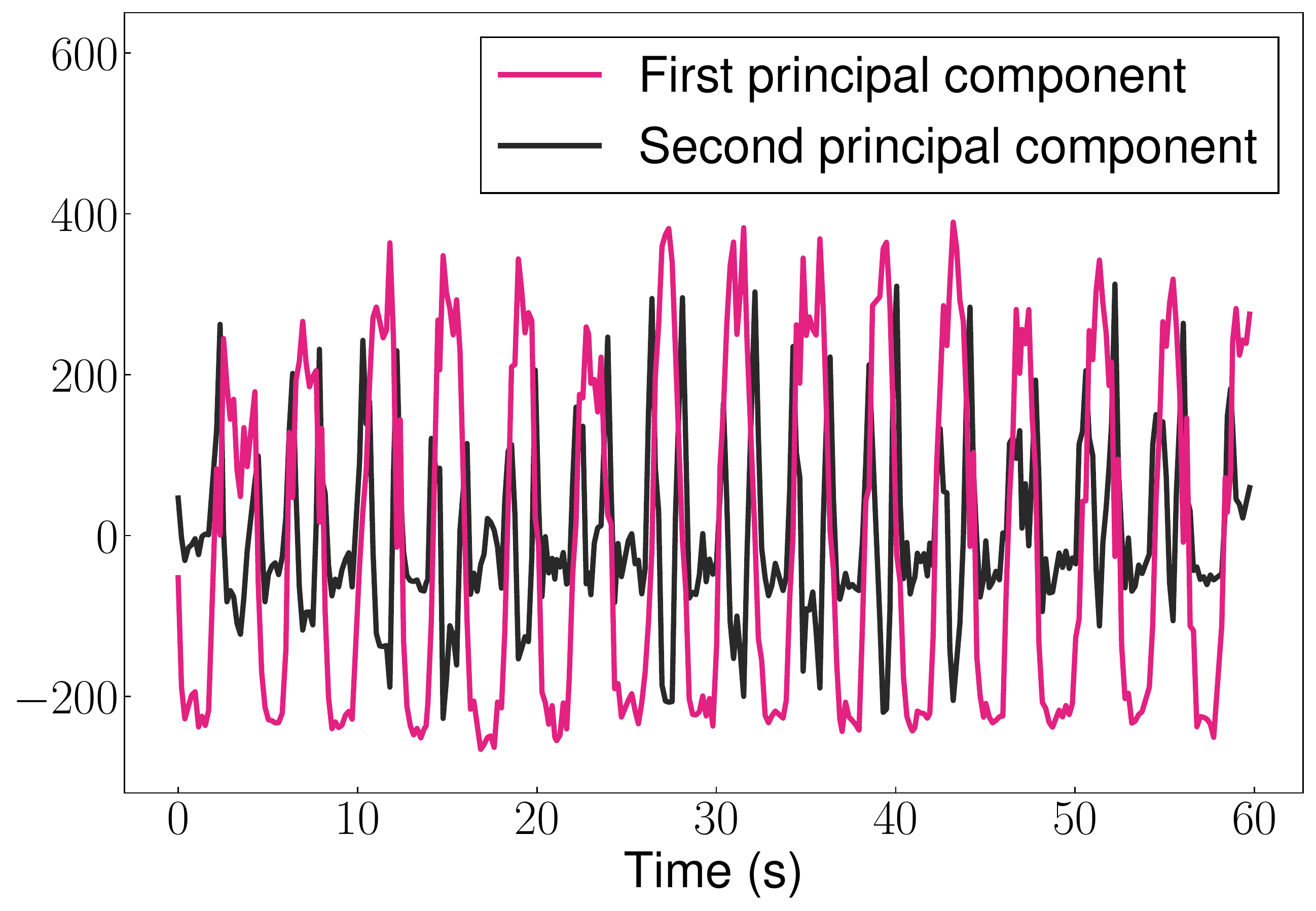}\label{subfig:after_preprocessing}} \\
    \subfloat[Contribution rate of principal components.]{\includegraphics[width=\columnwidth]{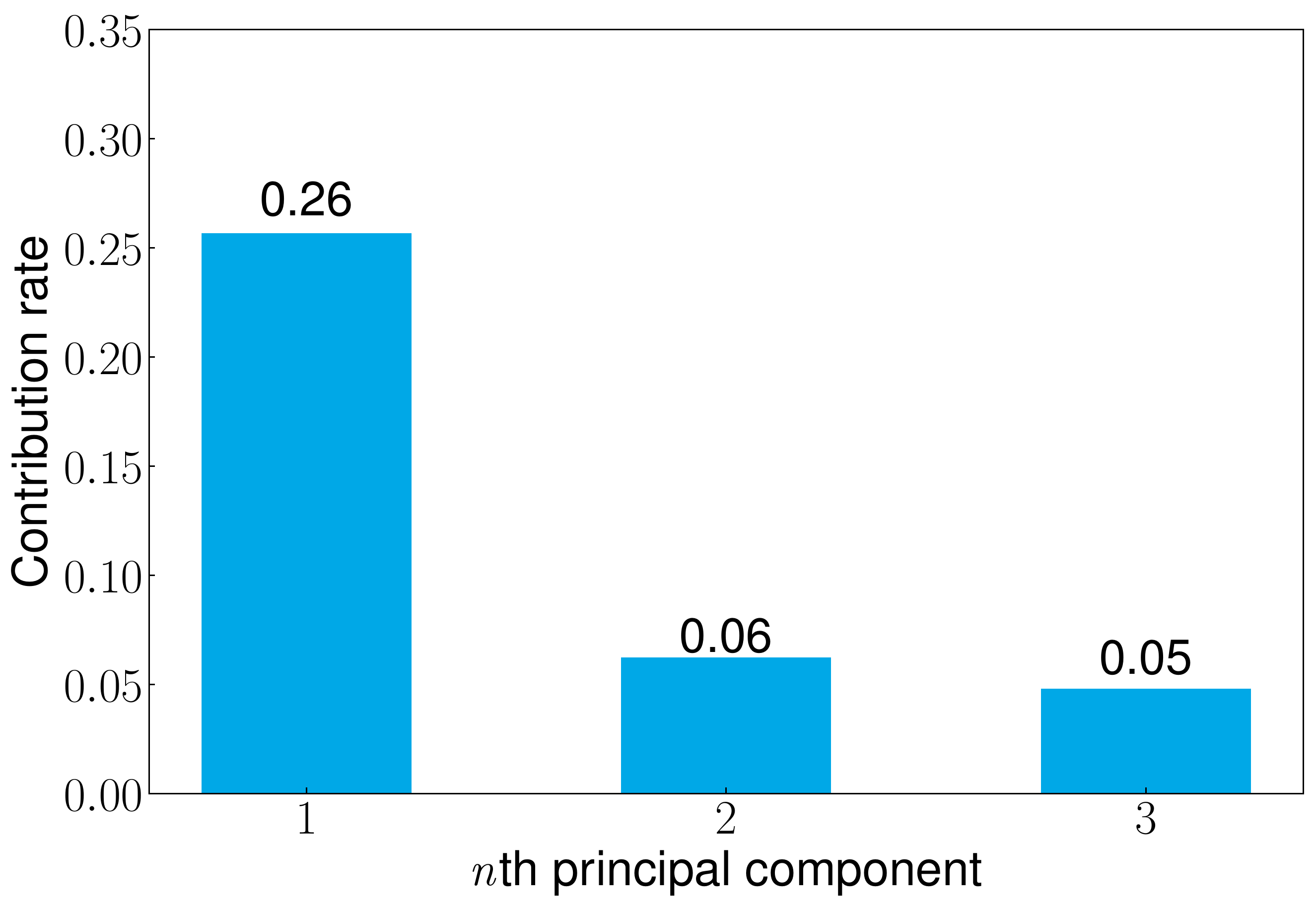}\label{subfig:contribution_rate}} \\
    \caption{
        Effectiveness of preprocessing.
        Periodic fluctuation owing to respiration becomes evident after
        PCA, where the first principal component exhibits a large contribution rate compared with other principal components.
    }
    \label{fig:preprocessing_results}
\end{figure}

\begin{figure}[t]
    \centering
    \includegraphics[width=\columnwidth]{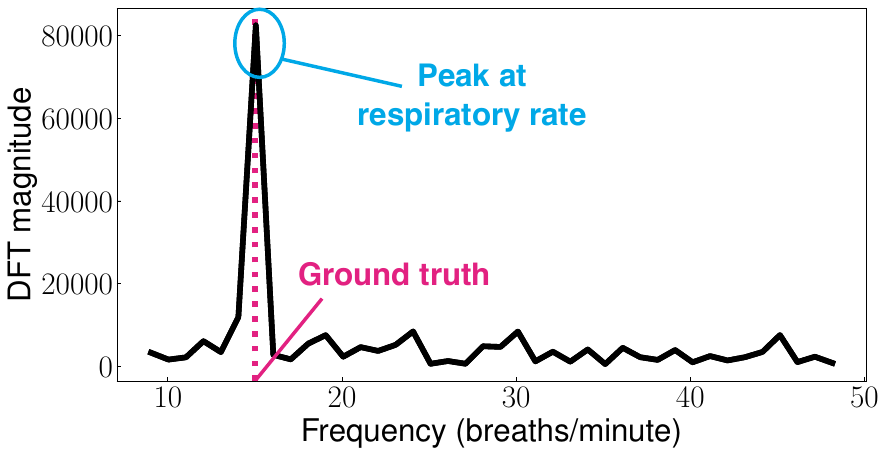}
    \caption{
        DFT output.
        The output exhibits
        one peak at a frequency corresponding
        to subject's ground-truth respiratory
        rate of $15\,\mathrm{breaths/minute}$.
    }
    \label{fig:dft_output}
\end{figure}

\begin{figure}[t]
    \centering
    \includegraphics[width=\columnwidth]{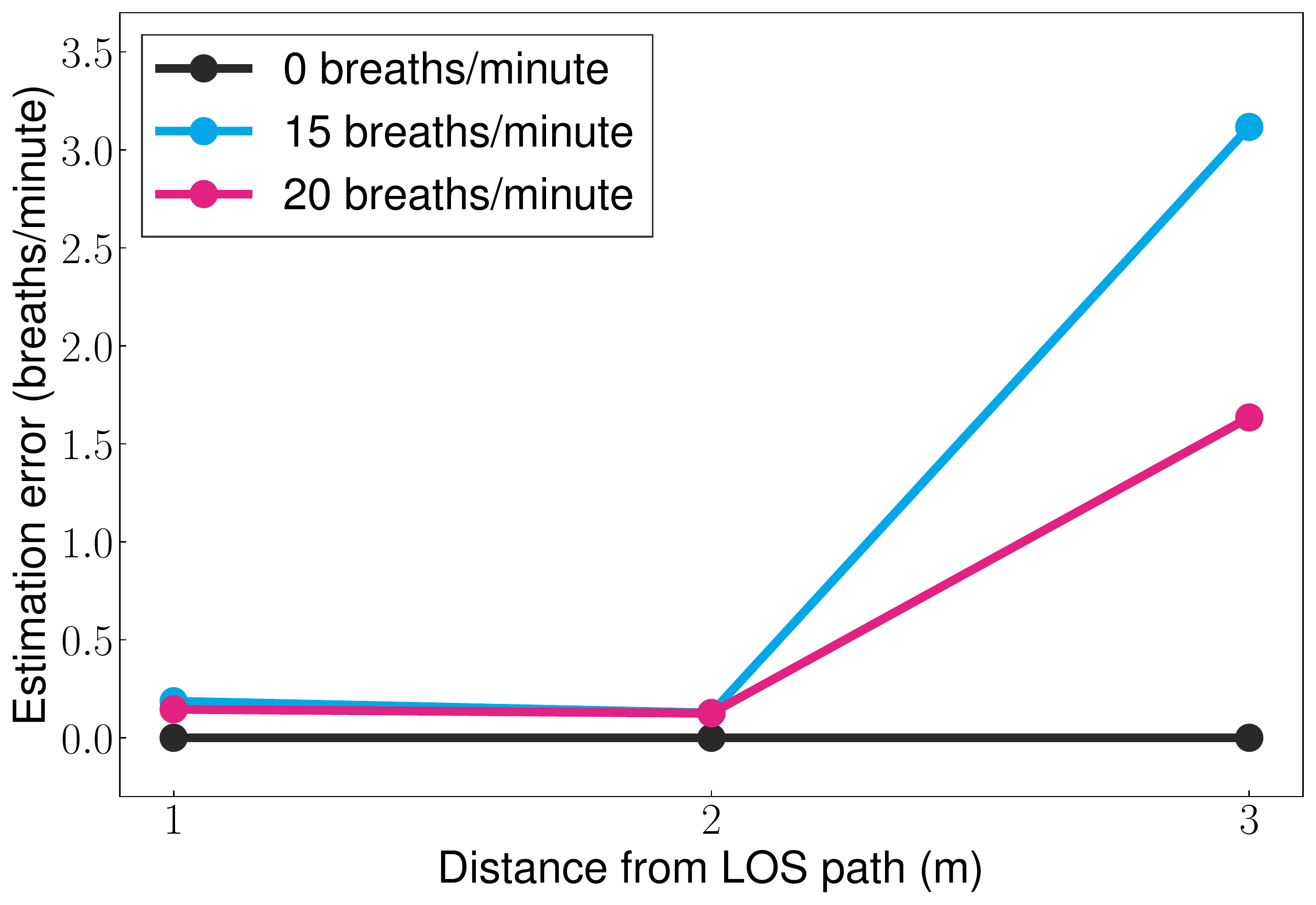}
    \caption{
        Estimation error as a function of the distance from the LOS path.
        The errors are lower than $3.2\,\mathrm{breaths/minute}$
        at all subject distances.
        When the distance is smaller than 2\,m,
        the estimation error is lower than
        $0.20\,\mathrm{breaths/minute}$.
    }
    \label{fig:estimation_error}
\end{figure}

\noindentbf{Preprocessing.}
\figurename~\ref{fig:preprocessing_results} confirms the effectiveness of preprocessing on the decompressed BFMs.
Figs.~\ref{fig:preprocessing_results}\subref{subfig:after_preprocessing}
and \ref{fig:preprocessing_results}\subref{subfig:contribution_rate} show the data after PCA and
the contribution rate of each principal component at
a distance of $1\,\mathrm{m}$ and a breathing rate $15\,\mathrm{breaths/minute}$.
Periodic fluctuation has become evident in the first principal component.
It has a large contribution rate
compared with other principal components.
The second and the third principal components possess smaller contribution rate and have been considered to represent
the fluctuation of the other channel information, e.g., noises.
\vspacelb
\noindentbf{DFT Output.}
\figurename~\ref{fig:dft_output} exhibits
a clear peak at the frequency of the ground-truth breathing rate of $15\,\mathrm{breaths/minute}$.
This figure shows the output of DFT performed on the preprocessed data
at $1\,\mathrm{m}$ and $15\,\mathrm{breaths/minute}$.
By extracting this peak, we can estimate the respiratory rate as discussed below.
\vspacelb
\noindentbf{Respiratory Rate Estimation.}
\figurename~\ref{fig:estimation_error} confirms the feasibility of
accurate respiratory rate estimation based on frame capturing.
The estimation errors are lower than $3.2\,\mathrm{breaths/minute}$ at all the distances from the LOS path.
When the subject distance is $1$ or $2\,\mathrm{m}$ from the LOS path, the estimation error is lower than
$0.20\,\mathrm{breaths/minute}$, which would be sufficiently accurate
for medical use cases referred in \sectionname~\ref{sec:introduction}.
This result reveals that respiratory rate can be estimated using BFMs in the frames captured
without firmware modifications nor dedicated WiFi chipsets.

However, the estimation error at a distance of $3\,\mathrm{m}$ is higher than those at
1 or 2\,m,
owing to the signal-to-noise power ratio pertaining to respiratory detection.
The received signal power reflected at the subject's chest
becomes much weaker than the direct signal along the LOS path.
The proposed method may have extracted
the fluctuations of channel information in LOS path
rather than the effect of breathing motion.

\section{Conclusion}
\label{sec:conclusion}

This paper presented a frame-capture-based respiratory rate estimation method.
We proposed an approach using digital signal processing.
Specifically, we applied PCA to isolate the effect of chest fluctuation from other channel information,
and performed DFT to extract respiratory rate.
The experimental results confirmed the feasibility of respiratory rate estimation based on frame capturing,
where estimation error is lower than $3.2\,\mathrm{breaths/minute}$.
This implies that the applicability of frame capturing can be extended to various sensing tasks.

\section*{Acknowledgment}
This work was supported in part by JSPS KAKENHI Grant Number JP18H01442, JP20H04156, and JP20K21793.

\bibliographystyle{IEEEtran}
\bibliography{bfmrespiration_ccnc2022}

\end{document}